# Gamed-based iSTART Practice:
# From MiBoard to Self-Explanation Showdown

Justin F. Brunelle[1], G. Tanner Jackson[2], Kyle Dempsey[2], Chutima Boonthum[3],
Irwin B. Levinstein[1], Danielle S. McNamara[2]

[1] Old Dominion University
[2] University of Memphis
[3] Hampton University

**Abstract**

MiBoard (Multiplayer Interactive Board Game) is an online, turn-based board game that was developed to assess the integration of game characteristics (point rewards, game-like interaction, and peer feedback) and how that might affect student engagement and learning efficacy. This online board game was designed to fit within the Extended Practice module of iSTART (Interactive Strategy Training for Active Reading and Thinking). Unfortunately, preliminary research shows that MiBoard actually reduces engagement and does not benefit the quality of student self-explanations when compared to the original Extended Practice module. Consequently the MiBoard framework has been revamped to create Self-Explanation Showdown, a faster-paced, less analytically oriented game that adds competition to the creation of self-explanations. Students are evaluated on the quality of their self-explanations using the same assessment algorithms from iSTART Extended Practice module (this includes both word-based and LSA-based assessments). The technical issues involved in development of MiBoard and Self-Explanation Showdown are described. The lessons learned from the MiBoard experience are also discussed in this paper.

## Introduction

Intelligent Tutoring Systems (ITSs) such as iSTART rely on repetition and consistency to enforce and provide opportunities to practice desired skill sets. Unfortunately, this process can become tedious and boring. Applying game components such as points and awards to a practice scenario has been shown to be effective in increasing engagement (Rowe, 2008). Games are considered a break from the traditional educational environment because they enforce the skill sets and knowledge bases provided by lecture (Gredler, 2004). However, the integration between game components and educational values in an intelligent tutoring system must be carefully balanced in order to increase user engagement while still providing ample opportunity to practice; one aspect should not out-weigh the other.

In the particular case of iSTART self-explanation training, MiBoard has been added with the aim to alleviate the tedium of the system and increase engagement among its users. Unfortunately, the balance between game components and training was uneven, which led to disengagement among users, and more poorly executed practicing. Also, the method with which the self-explanations were evaluated was inconsistent with the rest of the iSTART Practice. This paper gives a description of how MiBoard works as part of iSTART Practice and an account of its successes and failures. This paper also outlines the proposed replacement for MiBoard, Self-Explanation Showdown, which aims to correct the short comings of MiBoard while incorporating the iSTART assessment algorithms (McNamara, Boonthum, Levinstein, & Millis, 2007).

## Introduction to MiBoard and iSTART

MiBoard (Multiplayer Interactive Board Game) is the computerized game version of Rowe's physical version - iSTART: The Board Game (Rowe, 2008). iSTART is a web-based tutor for high school students to improve their reading and thinking skills that includes an extended practice component. Currently, students are guided through extended practice modules which provide repetitive practice using the iSTART strategies (see below *iSTART*) to create self-explanations. Students are given a text in which they are to create self-explanations for each of several targeted sentences. Multiple texts are given to the students with which to practice using the reading strategies.

Research with iSTART has indicated the need for students to have extended practice with reading strategies (O'Reilly et al., 2004). This is because the effects of the initial iSTART training tend to taper over time. Therefore, students need additional, extended practice after the initial training. Unfortunately, research has also indicated that iSTART, while relatively engaging for most students initially, can become tedious perhaps because its layout is somewhat static, or possibly because the interaction (during extended practice) is predictable. A decline in engagement over time may also result from the lack of explicit incentive for the students to achieve mastery of the reading strategies. Rowe's (2008) game showed potential for alleviating the tedium of iSTART Extended Practice. MiBoard addresses all of the above concerns, including the lack of engagement.

MiBoard is an extension of iSTART that allows students to practice the skills targeted by iSTART in a more engaging and stimulating environment, while collaborating with their peers in a more social and structured educational forum. MiBoard is a 3- or 4-player turn-taking board game that gives players practice in making and analyzing self-explanations of sentences that occur in the context of longer texts. The main goal of MiBoard is to observe the effects of introducing game components to a learning environment, with particular attention to effects on engagement.

The current extended practice emphasizes repetitively creating self-explanations with any of the strategies, while MiBoard emphasizes both analytically creating self-explanations using a single, targeted strategy at a time and identifying the use of various strategies in peer self-explanations.

Preliminary research has shown that MiBoard falls short of the goals set by the researchers. The interface leads to reduced user engagement, and the nature of MiBoard's game play differs more greatly than anticipated from iSTART Practice. A new game, Self-Explanation Showdown, is in development with the goals of correcting the failures in the MiBoard interface, improving engagement, and improving self-explanation quality.

## iSTART: The Board Game

iSTART: The Board Game (iSTART: The Board Game) was developed by Rowe (2008). iSTART: The Board Game was created to investigate the effects of converting the practice module of iSTART to a game-based adaptation.

While playing iSTART: The Board Game, students take turns reading a text and generating self-explanations. A player begins as the Reader, and generating a self-explanation using a reading strategy drawn from a card. After saying aloud the self-explanation, the other players, called Guessers, guess as to which strategy the Reader used in his/her self-explanation. Any disagreements about which strategy was used are resolved in a discussion round. After these steps, the Reader rolls a die to move the Reader's token, and draws an event card, which may move his/her piece forward or backward, similar to the Chance cards in Monopoly.

Rowe (2008) reported results indicating that iSTART: The Board Game was an effective form of practice but the game was not intended as a replacement for any of iSTART's existing modules. He also found that players found the game an enjoyable method of practicing with iSTART. Rowe hypothesized that a digital game would provide users with an alternative means to practice with their peers without being in the same physical location. Rowe concluded that this game could provide an effective tool for alleviating the monotony and tedium often associated with the long-term interaction that extended practice imposes upon participants.

## MiBoard

MiBoard (Multiplayer, Interactive Board-game) is the computerized game version of Rowe's physical board game mentioned in the above section. The game-play of MiBoard directly mimics the game-play of iSTART: The Board Game. As such, each of the stages of game-play in iSTART: The Board Game (reading, guessing, discussing, etc.) is represented as entities (screens) in MiBoard (Reader Screen, Strategy Identification Screen, etc.). This direct translation from physical system to virtual system was made for simplicity purposes and to keep the game play of MiBoard as close to iSTART: The Board Game as possible.

After the Reader finishes typing a self-explanation, the Guessers are shown the Reader's self-explanation on the Strategy Identification Screen. The Reader uses the same interface to verify that the specified strategy was used. Each Guesser decides which single strategy is most prominent in the self-explanation and defends this choice by constructing an argument with the help of a Cascading Menu Block. The Cascading Menu Block (CMB) is used to provide structure for the responses (Figure 1). The CMB also reminds the students of the meanings of the strategies and how to identify their use in a self-explanation. Students are allowed a more free form exhibition of their knowledge in the discussion that may ensue.

Free-form responses may have worked in the supervised game conducted in Rowe's experiment, but more structure is given in the unsupervised environment existing in MiBoard in the form of the CMB seen in Figure 1. Students must also identify the strategies used in their opponents' self-explanations while playing MiBoard. The CMB plays two roles. The first is to reinforce how a strategy is used (e.g., paraphrasing is restating the sentence in your own words). The intention is for the students to see the reasoning from the CMB and to realize how the strategy was either used or not used in the self-explanation. Second, the CMB allows students to provide accurate and detailed feedback to the reader in a distributed environment. The CMB is also used in MiBoard because the algorithm used in iSTART Practice is not appropriate for determining strategies used in self-explanations submitted with MiBoard; the iSTART algorithm is so far unable to pinpoint strategies used, and only can determine the overall quality of a self-explanation. For this reason, the players must use the CMB as a method of self-explanation evaluation instead of MiBoard automatically analyzing self-explanations with the iSTART algorithm.

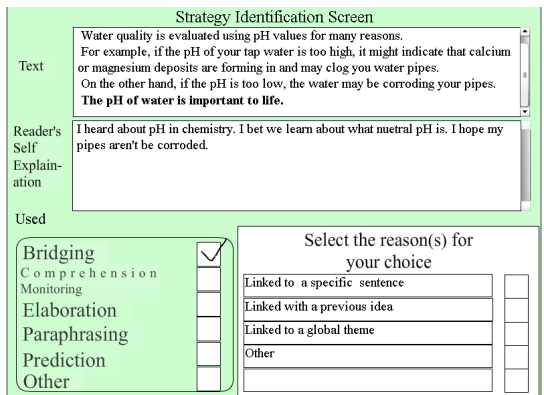

Figure 1. Strategy Identification Screen, where the player will identify the strategy used in the reader's self-explanation and then a cascading menu block is presented.

### Strategy Identification Screen

At the Strategy Identification Screen (Figure 1), players select the strategy they think was focused on by the Reader. The Guesser may only choose one such strategy at this stage in the game.

**Cascading Menu Block.** The Cascading Menu Block (Figure 2) is part of the Strategy Identification Screen. It is called cascading because each time a user clicks on a check box, a new set of options appears on the screen. A user is asked to click on a strategy followed by a reason for that selection (such as, *Linked to a specific sentence*), and then is asked to highlight the part of the self-explanation in which that particular strategy was used.

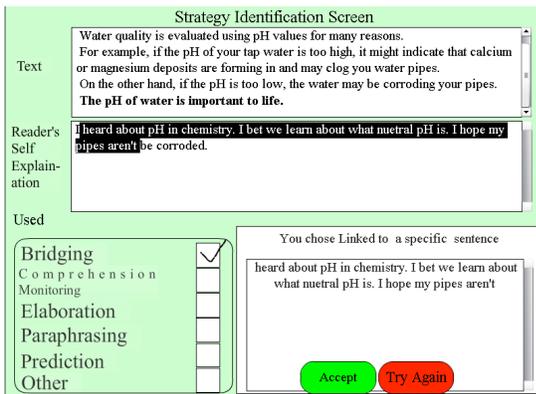

Figure 2. Guesser Cascading Menu Block. The player not only has to identify the strategy used, but also provide the reason why the player thinks the strategy was used.

### Preliminary Results with MiBoard

The development of MiBoard required solving numerous technical problems. MiBoard also contains a number of algorithms that serve as templates for future serious game endeavors. However, several problems arose with the user interface.

Preliminary tests using groups of students have been conducted with MiBoard. One test group was composed of freshmen college students, the other was composed of senior high school students. Both groups were given an abbreviated iSTART introduction to familiarize the students with the reading strategies and creation of self-explanations. They were then introduced to the rules and gameplay of MiBoard, and answered a series of questions. The students were then given a small window of time to play. The college freshmen played for roughly one round, and the high school seniors played for 20 minutes. Upon completion of the test, the students completed a questionnaire.

The preliminary test results have shown that MiBoard actually leads to a reduction in engagement due to three issues: points in which the interface may be stalled or the game-play particularly slow, unexpected changes in the interface, and the lack of a self-explanation evaluation algorithm.

**Lulls in Game Play.** The game has several periods in which players have nothing to do, and thus, become bored with the game and stop paying attention to the tasks to be completed. One such period is while the reader is creating a self-explanation. At this point, guessers do not have a task to keep them occupied. For this reason, the guessers lose interest in the game quickly. In iSTART: The Board Game, Guessers were able to see the reader compose the self-explanation; MiBoard has no method of showing the Guessers this process. MiBoard players simply must wait (for longer than desired) for the end result to suddenly appear. Although MiBoard provided a chat for the unoccupied players, they did not make use of it.

The length of time spent composing self-explanations also leads to too little time being spent emphasizing the components of the game that made it fun to play. Too little time is spent rolling the dice and using power cards to move on the board. These game aspects arouse competition and provide the fun aspects of MiBoard. There is also a very weak link between the game components and the purpose of MiBoard; that is, there is no correlation between how well the user self explains and how many spaces across the board the user moves. Engagement was reduced due to the limited exposure to game components and due to the lengthy pauses in game play.

**Screen Progression.** Another significant issue in MiBoard is that the game's screen progression is confusing. The game jumps from screen to screen, and this sequence can confuse the user easily; there is an uneven flow to the game. The player experiences a large amount of time in which there are no tasks to be accomplished (such as during the Reader's composition of his self-explanation). This leaves players sitting at the Main Board Screen, but without the ability to interact. When it is time for the player to accomplish a task, a new screen suddenly appears and requires user input. This apparently illogical and sudden change of screens was a source of confusion. This disjoint and unintuitive flow causes users to become disengaged.

**Self-Explanation Evaluation Algorithm.** MiBoard relies on player participation to score a reader on the quality of self-explanations. Guessers identify which strategy has been used in a self-explanation. Because the LSA and word matching algorithms used in iSTART do not provide specific information on which strategy is used, the algorithm is unsuitable for MiBoard. Also, automatic evaluation of self-explanations was not a project objective. If needed, a modification of current iSTART evaluation has to be made so that it gives strategy usages rather than the overall quality; this would be a major undertaking and would be outside the canon of MiBoard.

Further, iSTART Training emphasizes strategy identification as a secondary activity, and only emphasizes single strategy identification in their own self-explanations. For this reason, students are unfamiliar with identification of strategy use in other self-explanations augmented with providing evidence of the strategy usage. Also, moving through the CMB was time consuming, and students in the test groups completed the CMB at varying rates, further amplifying the lulls in gameplay.

**User Feedback.** During the post-test questionnaire, players gave feedback that reinforced the problems mentioned in this paper. The following are verbatim responses to the question "MiBoard would be better if: " from actual users:

> "it was easier to understand how to play. People like simple things. They cant comprehend things that are out of their league such as this game. Maybe make the game less complicated"

> "the kinks need to be worked out like the room capacity, the pace of the game, and the excitement of the overall experience. "

Of the 24 high school seniors participating in one of the two tests, only 9 completed the post-test questionnaire. The remainder were so confused and disengaged with the game that they failed to answer the follow-up questions. All 9 responses reflected negative experiences with MiBoard. Further, during actual gameplay, the pace of the game was so slow that only one game was able to finish a round (all 4 players produced self-explanations) of the 6 games occurring during the 15 minutes allotted for game play. For this reason, there are very few self-explanations available for thorough analysis. The following is a player self-explanation from the high-school student testing that is representative of the few self-explanations collected:

> "I don't have a clue"

The self-explanations entered by players further express their unwillingness to remain engaged with MiBoard. The few self-explanations collected shows that the game is far too slow to be an effective practice tool.

## Self-Explanation Showdown

Self-Explanation Showdown is to be an improved multiplayer flash game to replace MiBoard. Self-Explanation Showdown provides a more pure form of practice than MiBoard, and will be simplified to increase user understandability and engagement. Further, Self-Explanation Showdown provides a single and multiplayer version. This provides an additional edge over MiBoard; single player games often provide drills on the knowledge base (Knowledge Adventure, 1999), while multiplayer games allow users to understand the inappropriate versus appropriate scenarios in which to use a skill set, and also identify the use of skills through peer examples (Rowe, 2008).

Self-Explanation Showdown was built on the same technical framework as MiBoard .. It is a two-player game in which users read a text, including a target sentence. The players create self-explanations for the target sentences. Each user's self-explanation is scored by the iSTART algorithm which rates its overall quality. The user with the highest self-explanation score from the algorithm wins the point. Ties are resolved by an additional round worth 2 points (see Figure 3).

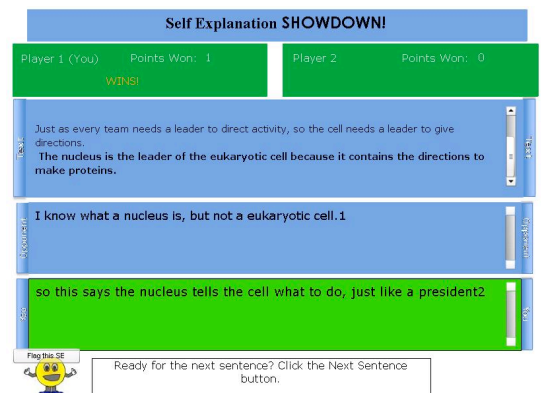

Figure 3. Self-Explanation Showdown interface shows the self-explanations in a single, simple window.

**Lulls in Game Play.** Self-Explanation Showdown either eliminates, or drastically reduces, the phases during MiBoard in which players have no task to accomplish, such as during self-explanation generation and during guessing. Both users create self-explanations at once, eliminating a single user being without a task to complete. Also, the time a player may have to wait for an opponent to *finish* a self-explanation in Showdown is far less than the amount of time a Guesser in MiBoard has to wait for the Reader to compose an entire self-explanation. The iSTART algorithm is used instead of user guessing, eliminating the delay occurred when waiting on a player vote. The reduction of wait-times in both instances should make the delays more bearable in Showdown. Further, during system testing, the development team noticed a drastic reduction in wait time during game play.

**Screen Progression.** Self-Explanation Showdown consists of a single screen. This simplifies the interface and reduces user confusion. All state progressions within the game are achieved with predictable timers and all progressions are user controlled; all timers used in Self Explanation Showdown use intervals of 2 seconds whereas, in MiBoard, no timer was used and players were confused by the apparent sudden appearance of screens.

**Self-Explanation Evaluation Algorithm.** The iSTART algorithm used in practice and extended practice modules is used directly in Self-Explanation Showdown. The self-explanation as well as all pertinent information (such as text number and sentence number) is passed to the algorithm. The algorithm returns a quality of self-explanation in a 4-point scale to the game: 0 (Unsatisfactory: irrelevant, too short, or too similar), 1 (OK), 2 (Good), and 3 (Excellent). It is worth noting that only a quality score is used in the current implementation of the Self-Explanation Showdown.

## Technical Aspects and Innovations

MiBoard and Self-Explanation Showdown were created using the Flash programming language ActionScript 3.0, JavaScript, Java Server Pages (JSP), AJAX, MySQL, and ElectroServer. The framework on which MiBoard is built is an innovative system of algorithms and languages. The ability of the MiBoard framework to interface dynamically with a database through ActionScript is of particular significance. The use of the ExternalInterface object to interface between movies on the same client, distributed movies, and a database is the most important aspect of the MiBoard framework. The use of these languages together is an innovative way of linking multiple languages and tools together, utilizing each of their unique strengths to accomplish a single, seamless system.

**ElectroServer.** ElectroServer is a multiplayer server product that facilitates interaction between many connected users and can be used for real-time audio and video streaming and recording. It is particularly useful for hosting Flash games. With its multiplayer feature, ElectroServer is suitable to be used to handle the multiplayer game in iSTART. ElectroServer works by allowing client applications, such as Flash, Java, or Silverlight, to connect to it via socket and log in. While connected, the server can push data to the client or the client can make requests of the server at any time. ElectroServer specializes in allowing communication between Flash movies through its own set of abstract data types (ADTs) and code structure.

Rooms and zones are ADTs in ElectroServer that proved particularly useful in the development of MiBoard. A room is a collection of users playing a single game. These users can easily communicate to achieve chatting or multiplayer game play. A zone is a collection of rooms. Chatting can occur as public messages sent to an entire room of users, or private messages that are sent to one or more specific users in any room. These public messages are also used in game state synchronization.

In transforming Rowe's (2008) game, a single room is used to represent a board of three or four players. The room enables each player to be synchronized and supports chatting among the players. Because MiBoard only supports 3 to 4 players, maximum capacities are set on each room. MiBoard automatically searches for the first room that has not yet started and is not yet full. If there are no rooms satisfying these criteria, a new room is created and the user enters that room. Once a room has three players, the game may begin. Beginning a game prevents any further players from entering that room.

**Passing Control in MiBoard.** MiBoard utilizes a round-robin master-slave relationship among participating clients. Each client contains the code to run the game in its entirety. When the client is a Reader, the client controls each of the other connected clients by passing public messages to each client. The clients receive the messages and parse them to determine the desired action. These messages are sent from the chat to each client connected to the game. Upon completion of the Reader's turn, the control is passed to the next player, making that client the master, and reverting the previous master to a slave.

This method of control passing gives the MiBoard framework flexibility in handling player attrition during a game. MiBoard was originally intended to be developed further, which would include handling user logouts or internet connection failures. Using this round-robin master-slave relationship, a player could leave a round of MiBoard, and the game could proceed with one less player, but otherwise unchanged. Thus, experimental data would not be lost upon the occurrence of a network failure, as is possible in a distributed environment.

**Synchronizing Information.** String parsing and recognition of user vs. game communication is essential. The messages passed have a very strict format, and cause the game to behave properly; the messages synchronize the game at each computer at which a user is playing the game. Codes are inserted into the messages for game play, and messages without codes are messages sent by the users.

**ActionScript.** The underlying infrastructure of MiBoard is particularly interesting. ActionScript 3.0 is not made to communicate with databases or other exterior entities. ActionScript only references its calling entity via the object ExternalInterface. ExternalInterface has a property "call" which tells the calling entity to invoke its function specified in the call. For example, ExternalInterface.call( "myFunc", "myParam" ) invokes the calling entity's myFunc function with the parameter myParam. Because MiBoard consists of a chat movie and a game movie imbedded in a JSP page containing JavaScript, MiBoard is able to call functions in JavaScript through the ExternalInterface object. Two movies are separate entities embedded in the same JSP page. Thus, one movie interacts with the other by calling functions in JavaScript that call functions in the other movie. When the board movie wants to tell all connected players that a player has moved 2 spaces forward, the board movie tells the JavaScript to tell the chat that the player just moved 2 spaces. The chat broadcasts that message to all connected players.

Java Server Pages (JSPs) and JavaScript. JSP and JavaScript are chosen for MiBoard for a couple of reasons. First, all iSTART modules are developed in JSPs. JavaScript allows JSP pages to interact with the users, communicate back to the server, and store records in the databases. Second, to minimize the integration between iSTART and MiBoard, same platforms should be used. Although the game components are developed in Flash, ActionScript's ExternalInterface class allows communication between ActionScript/Flash and JSP/JavaScript.

**MySQL.** Logging the progress of players is essential in analyzing the effectiveness of MiBoard. This logging is done in a MySQL database. Because communication with the MySQL database occurs in iSTART's JSP pages, MiBoard must communicate within the JSP pages. JSP is a server-side language, and therefore cannot interface with the database after the page has been rendered and loaded. The method used to circumnavigate that obstacle involves the aforementioned strategy. ActionScript tells the JavaScript that it would like to log data (which is passed as a parameter to the JavaScript function). The JavaScript parses the data, and creates an iFrame of length and width 0. This invisible iFrame contains a new JSP page, which takes a MySQL query as a parameter. This new JSP page executes the passed query, and closes.

Alternatively, AJAX is used to retrieve message from a database. The method is the same as described above, except after the JSP page has queried the database, it writes (in XML) the information retrieved from the database on the HTML page produced by the JSP. This HTML page is returned to the JavaScript, which, in turn, parses the data using the XML tags. This functionality is not used in MiBoard, but is a necessary part of the framework since Self-Explanation Showdown requires the ability to query a database during game play.

## Conclusion & Future Work

Rowe's (2008) physical board game was converted into a virtual game. This conversion accomplished several technical goals, including new methods to retrieve experiment data from a Flash movie. The techniques used to develop MiBoard also provide the methodologies for future games, thus providing the basis for a general framework for Flash game development.

However, the virtual game is not as engaging and preliminary tests show it may actually reduce user engagement. This is possibly due to the inability to effectively transform the game components of iSTART: The Board Game into a virtual environment while retaining the maximum fidelity of the original game play and layout.

Self-Explanation Showdown is a more pure form of practice and aims to provide a more effective method for students to use the iSTART strategies in an engaging and educational environment. It is currently being developed using the MiBoard framework and is intended to continue the technical, as well as experimental, work of MiBoard. The framework allows Self Explanation Showdown to pass public messages between client machines, and allows the iSTART algorithm to be incorporated into the game via AJAX and JavaScript function calls.

Future work with the iSTART project will focus on improving the human-computer interface of Self-Explanation Showdown to further increase user engagement. Self-Explanation Showdown will also provide essential feedback and data collection using the iSTART algorithm.

## References

Chi, M.T.H., Bassok, M., Lewis, M., Reimann, P., and Glaser, R. 1989. Self-explanation: How students study and use examples in learning to solve problems. *Cognitive Science*, *13*, 145-182.

Chi, M.T.H., De Leeuw, N., Chiu, M., and LaVancher, C. 1994. Eliciting self explanations improves understanding. *Cognitive Science*, *18*, 439-477.

Gredler, M.E. 2004 Games and simulations and their relationships to learning. *D.H. Jonassen (Ed.), Handbook of Research for Educational Communications and Technology: A Project of the Association for Educational Communications and Technology (2nd ed., pp.571–582)*. Mahwah, NJ: Lawrence Erlbaum Associates.

Knowledge Adventure. 1999. *Math Blaster [video game]*. Los Angeles: Knowledge Adventure, Inc.

Landauer, T.K., McNamara, D.S., Dennis, S., and Kintsch, W. eds. 2007. *Handbook of Latent Semantic Analysis*. Lawrence Erlbaum, Mahwah, NJ.

McNamara, D.S., Levinstein, I.B. and Boonthum, C. 2004. iSTART: Interactive strategy trainer for active reading and thinking. *BRMIC*, *36*, 222-233.

McNamara, D.S., Boonthum, C., Levinstein, I.B., and Millis, K.K. 2007. Evaluating self-explanation in iSTART: Comparing word-based LSA systems. T. Landauer, D.S. McNamara, S. Dennis, and W. Kintsch eds., *Handbook of Latent Semantic Analysis* (pp. 227-241). Lawrence Erlbaum, Mahwah, NJ.

O'Reilly, T. P., Sinclair, G. P., and McNamara, D. S. 2004. iSTART: a web-based reading strategy intervention that improves students' science comprehension. Kinshuk, D.G. Sampson, and P. Isaias eds., *Proceedings of the IADIS-2004, Lisbon, Portugal.* IADIS Press, pp. 173-180.

Rowe, M. 2008. Alternate forms of reading comprehension strategy practice and game-based practice methods. *Doctoral Dissertation*, Psychology Department, the University of Memphis.